\documentclass[prl,twocolumn,showpacs,preprintnumbers,amsmath,amssymb]{revtex4}
\include{epsf}

\begin{document}
\title{An all silicon quantum computer}
\author{T. D. Ladd}
\email{tladd@stanford.edu}
\author{J. R. Goldman}
\author{F. Yamaguchi}
\author{Y. Yamamoto}
\altaffiliation[Also at ]{NTT Basic Research Laboratories, 3-1
Morinosato-Wakamiya Atsugi, Kanagawa, 243-0198, Japan.}
\affiliation{Quantum Entanglement Project, ICORP, JST, Edward L.
    Ginzton Laboratory, Stanford University, Stanford, California
    94305-4085, USA}
\author{E. Abe}
\author{K. M. Itoh}
\altaffiliation[Also at ]{PRESTO-JST.}
    \affiliation{Department of Applied Physics and Physico-Informatics,
    Keio University, Yokohama, 223-8522, Japan}
\date{\today}
\begin{abstract}
A solid-state implementation of a quantum computer composed
entirely of silicon is proposed.  Qubits are $^{29}$Si nuclear
spins arranged as chains in a $^{28}$Si (spin-0) matrix with
Larmor frequencies separated by a large magnetic field gradient.
No impurity dopants or electrical contacts are needed.
Initialization is accomplished by optical pumping, algorithmic
cooling, and pseudo-pure state techniques.
Magnetic resonance force microscopy is used for readout. This
proposal takes advantage of many of the successful aspects of
solution NMR quantum computation, including ensemble measurement,
RF control, and long decoherence times, but it allows for more
qubits and improved initialization.
\end{abstract}

\maketitle
%


The primary difficulty in the construction of quantum computers is
the need to isolate the qubits from the environment to prevent
decoherence, while still allowing initialization, control, and
measurement.  To date, the most successful experimental
realizations of multi-qubit, many-gate quantum computers have used
room-temperature, liquid NMR with ``pseudo-pure" states
\cite{nmrqc}.
These computers are able to maintain isolation from the control
and measurement circuitry by employing weak measurement on a large
ensemble of ${\sim}10^{18}$ uncoupled, identical molecules.
Although such a large, highly mixed ensemble may bring the
existence of entanglement into question \cite{braunstein99}, the
arbitrary unitary evolution afforded by the RF-controlled quantum
gates assures that these computers behave non-classically
\cite{schack1999}.  Their principal limitation results from their
small initial nuclear polarization. The size of the effective
sub-ensemble of nuclei contributing to the pseudo-pure state, and
hence the effective Signal-to-Noise Ratio (SNR), decreases
exponentially with each added qubit, leaving this method unlikely
to exceed the 10-qubit level without substantial modification
\cite{warren97}.


The proposals of Kane \cite{kane98} and others to use single
nuclear spins in a low-temperature solid solve the scalability
problem of solution NMR, but they introduce the problem of
single-nuclear-spin measurement.  It remains an experimental
challenge to fabricate a structure in which individual nuclei are
sufficiently coupled to an electronic system for single-spin
measurement, but also sufficiently isolated for long coherence
times.

\begin{figure}
\begin{center}
\epsfysize=3in \epsfbox{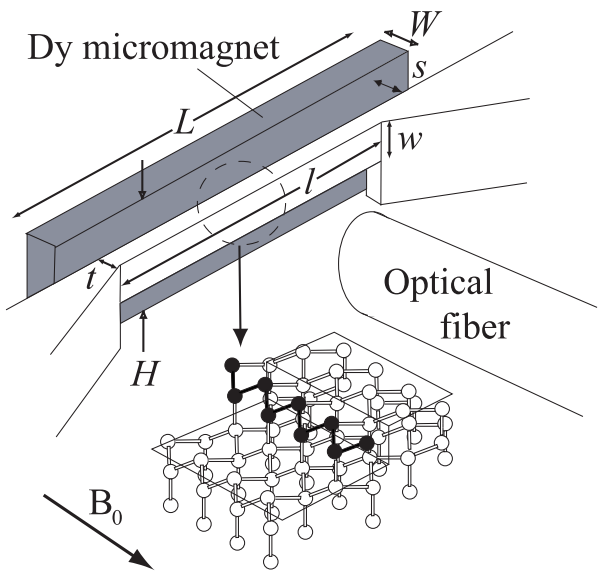}
\end{center}
\caption{The figure shows the integrated micromagnet and bridge
structure. The bridge has length $l=300 \ \mu$m, width $w= 4 \
\mu$m, and thickness $t= 0.25 \ \mu$m. The micromagnet has length
$L = 400 \mu$m, width $W = 4 \ \mu$m, and height $H = 10 \ \mu$m,
and produces a field gradient of $\partial B^z/\partial z = 1.4 \
T/\mu$m, uniform over a 100~$\mu$m by 0.2~$\mu$m region inside the
bridge. The insert shows the structure of the silicon matrix and
the terrace edge. The darkened spheres represent the $^{29}$Si
nuclei, which preferentially bind at the edge of the Si step.
\label{structure}}
\end{figure}

In this Letter, we propose a different solid-state NMR
implementation of quantum computation which introduces
electron-mediated cooling, but maintains the weak ensemble
measurement that has made solution NMR quantum computers so
successful.  The device is made entirely of silicon, with no
electrical gates or impurities.  As will be discussed below, the
qubits are spin-1/2 nuclei that are located in relatively isolated
atomic chains, 
as shown in Fig.~\ref{structure}. The nuclei within each chain are
distinguished by a large magnetic field gradient created with a
nearby microfabricated ferromagnet \cite{goldman2000}. Each
nucleus has about $10^5$ ensemble copies in a plane orthogonal to
its chain. This structure is embedded in a thin bridge whose
oscillations provide readout \textit{via} magnetic resonance force
microscopy (MRFM) \cite{sidles91,RYS92}.

The advantages of an all silicon implementation are many.
Foremost, the crystal growth and processing technology for silicon
are highly matured. In particular, the most sensitive structures
for force detection have been made from pure silicon. Also, it is
good fortune that the family of stable nuclear isotopes of silicon
is quite simple: 95.33\% of natural silicon is $^{28}$Si or
$^{30}$Si, which are both spin-0, and 4.67\% is $^{29}$Si, which
is spin-1/2, perfect for the qubit.  Thus, silicon is well suited
for nuclear-spin isotope engineering \cite{itoh2001}. Another
crucial motivation for the choice of silicon is the observation
that nuclei in a semiconductor may be polarized by
cross-relaxation with optically excited, spin-polarized conduction
electrons \cite{lampel68}. Although there are many other means for
dynamic nuclear polarization, optical pumping in semiconductors
has one important feature: the electrons whose hyperfine couplings
make the polarization possible recombine shortly thereafter and,
hence, do not contribute to decoherence during the computation.
The absence of any spurious spins, nuclear or electronic, should
leave the $^{29}$Si nuclei well decoupled from their environment.
%


In the following, we describe an example procedure for fabricating
the structures of Fig. \ref{structure} \cite{fabnote}. We start
with an isotopically depleted $^{28}$Si(111) ($<1\%~^{29}$Si)
wafer which has been miscut $1^{\circ}$ towards
$(\bar{1}\bar{1}2)$. Oxygen atoms are ion-implanted from the
surface of the substrate followed by appropriate annealing in
order to form a buried oxide layer, which will be removed later to
form an open space below the vibrating bridge.  The wafer is
transferred to a multi-chamber molecular beam epitaxy (MBE)
machine equipped with a scanning tunneling microscope (STM) and a
pre-heating stage.  It has been demonstrated that highly regular
arrays of steps can be produced on vicinal Si(111)7$\times$7 using
a simple multi-step annealing sequence \cite{sisteps}.  Due to the
high energy of interrupting or misplacing 7$\times$7 domains, this
procedure leads to atomically straight step-edges along the
$(1\bar{1}0)$ direction for up to $2\times 10^4$ lattice sites.
The average terrace width is $\sim$15~nm.
Once we verify the straight step edges using the STM, we transfer
the wafer to the growth chamber to form atomic chains of $^{29}$Si
using the step-flow mode at temperatures $T\sim 850^\circ$C.
Arrival of $^{29}$Si isotopes at the substrate surface induces a
surface transition from 7$\times$7 to 1$\times$1 \cite{laks89},
\textit{i.e.}, $^{29}$Si isotopes travel to the step edges and
form atomically straight lines.  We terminate the evaporation of
$^{29}$Si when the atomic chains are one atom wide and run
completely along the step. A $^{28}$Si capping layer of 15~nm is
grown on top of $^{29}$Si chains and we repeat the same multi-step
annealing, $^{29}$Si chain growth, and capping sequence to produce
replicas of parallel $^{29}$Si chains. The last step in the growth
process is the capping of ensembles of $^{29}$Si chains by a thick
$^{28}$Si layer.


Once formation of the $^{29}$Si wire block is complete, the narrow
bridge
is created with e-beam lithography. A plasma etcher removes
unprotected silicon down to the buried oxide layer. An HF vapor
etch or acid solution removes the exposed oxide to release the
structure; it is then dried in a critical point dryer. After
etching 3 $\mu$m further into the substrate and through the oxide
layer, the dysprosium (Dy) micromagnet can be evaporatively
deposited $s=2.1 \ \mu$m from the bridge and defined as a
parallelopiped using a shadow mask.


Calculations similar to those in Ref. \cite{goldman2000} show that
the magnetic field gradient due to the micromagnet is $\partial
B^z/\partial z= 1.4$ T/$\mu$m, which is persistent over the
thickness of the bridge and is superposed with a large homogeneous
field $B_0$ of  $\sim 7$~T. The distance in the $z$-direction
between two $^{29}$Si nuclei in an atomic chain, which we notate
$a$, is 1.9~\AA, so the gradient leads to a qubit-qubit frequency
difference of $\Delta\omega=a\gamma\partial B_z/\partial z=
2\pi\times 2$~kHz. The active region is a 100 $\mu$m by 0.2 $\mu$m
area in the center of the bridge, containing $N = 10^5$ chains
persisting over the bridge thickness. This active region is
sufficiently small and the magnetic field sufficiently homogeneous
that all equivalent qubits in an atomic plane lie within a
bandwidth of 0.6~kHz.


For initialization of the quantum computer, we propose to employ
optical pumping, algorithmic cooling, and pseudo-pure state
techniques. The premise of optical pumping is that nuclei exchange
Z\'eeman energy with a bath of electrons which have been
preferentially excited into a single spin state by circularly
polarized light.   The nuclei thereby relax thermally to an
effective spin temperature corresponding to the non-equilibrium
electron-spin polarization. Once those electrons recombine, the
nuclei retain their spin polarization for the ``dark" $T_1$ time,
which is extremely long (200 hours in Ref. \cite{lampel68}).  In
low-field ($\sim 1$~G) experiments at 77 K
\cite{lampel68,bagraev}, nuclear polarizations have not exceeded
0.1\% due to limited electron spin polarizations and long
recombination times, a result of silicon's indirect bandgap.
Improved nuclear polarization in silicon may be observable in
higher magnetic fields ($\sim$10~T) and lower temperatures
($\sim$1~K), and in silicon nanostructures where rapid
recombination of electrons \textit{via} surface states can help
maintain the electron spin polarization \cite{dal98}.

The physical cooling afforded by optical pumping is well suited to
complement an algorithmic cooling technique introduced by Schulman
and Vazirani \cite{sv,bmrvv2001}. This technique redistributes the
entropy among a register of qubits to a subregister that is then
discarded (decoupled and ignored) \cite{bmrvv2001note}. This
method is expensive; it takes time to perform the (classical)
logic operations and, worse, it sacrifices many qubits.  An
initial register of size $n_0$ will, for small initial
polarization $p_0$, shrink to size $n_0 p_0^2/2\ln 2$ if the
procedure is taken to the entropy limit. However, the very long
$T_1$ affords ample time for the procedure, and the number of
available qubits in our configuration before initialization can be
thousands. Moreover, the algorithm need not be taken to the
entropy limit, since large but still incomplete polarizations can
be handled with pseudo-pure state techniques, the consequences of
which will be discussed below in the context of SNR.


The secular component of the dipolar Hamiltonian which couples the
$i$th spin to the $j$th spin within one chain is written
\cite{abragam}
    \begin{eqnarray}\nonumber
    \hat{\cal H}_{ij} &=&\frac{\mu_0}{4\pi}\gamma^2\hbar^2
    \frac{1-3\cos^2\theta_{ij}}{r_{ij}^3}\hat{I}_{i}^z \hat{I}_{j}^z
    \\ &\equiv& -\hbar\delta\omega_{ij}\hat{I}_{i}^z \hat{I}_{j}^z,
    \label{ijcouple}\end{eqnarray}
where $r_{ij}$ is the length of the vector connecting the spins
and $\theta_{ij}$ is its angle with the applied field. Nearest
neighbors along the proposed atomic chains are not exactly
parallel to $(1\bar{1}0)$, but rather zig-zag with angle
$\theta_{i,i+1}$ satisfying $\cos^2\theta_{i,i+1}=2/3$, leaving
$\delta\omega\equiv\delta\omega_{i,i+1}=2\pi \times 0.4$~kHz.
Other terms of the dipolar Hamiltonian require the exchange of
energy in the amount of $\hbar\Delta\omega$ or more.  The long
$T_1$ in silicon indicates the inefficiency by which this energy
may be exchanged with degrees of freedom external to the nuclei.
In strongly coupled dipolar systems, this energy can be
compensated for by the dipolar bath of identical
nuclei\cite{gr73}.  In the present scheme, the members of the
nuclear ensemble are so far removed that this energy bath is
absent.  Hence, we assume these nonsecular terms are well
suppressed.  The Hamiltonian of Eq. (\ref{ijcouple}) may be
``switched off" by applying a periodic succession of narrow band
$\pi$ pulses at, for instance, the $i$th resonant frequency
\cite{haeberlin}. Simultaneous decoupling of more than two qubits
may be accomplished by timing the selective $\pi$ pulses according
to the entries of an appropriately sized Hadamard matrix; a pair
of qubits may be selectively recoupled in order to implement
two-bit gates \cite{leung99}.  Note that this Hadamard pulse
scheme serves the second purpose of refocussing inhomogeneous
broadening caused by the in-plane nonuniformity of the field
gradient and any small bulk susceptibility effects.


The presence of a large magnetic field gradient provides a natural
means for performing MRFM on a magnetization $M^z$, since this
technique is sensitive to the gradient force given by
$F^z=M^z\partial B^z/\partial z$. The experiment is performed in
high vacuum ($<10^{-5}$~torr) and at low temperatures (4~K). A
coil is used to generate the RF pulses for logic operations and
decoupling sequences; it also generates the continuous-wave
radiation for readout. An optical fiber-based displacement sensor
is used to monitor deflection of the bridge using interferometry.
Sub-\AA{ngstrom} oscillations can be detected; larger oscillations
can be damped with active feedback which avoids additional
broadening while maintaining high sensitivity~\cite{durig97}.

Readout is performed using cyclic adiabatic inversion
\cite{abragam}, which modulates the magnetization of a plane of
nuclei at a frequency near or on resonance with the bridge. The
spins of resonant frequency $\omega_i$ are irradiated with the RF
field $B^{x} =
2B_{1}\cos\{\omega_{i}t-(\Omega/\omega_m)\cos(\omega_{m}t)\}$,
where $\omega_{m}$ is the modulation frequency chosen to be near
the resonance of bridge oscillations, and $\Omega$ is the
frequency excursion, which should be much smaller than
$\Delta\omega$ \cite{RYS92}.  The $z$ component of the $i$th
plane's magnetization is deduced from the phase of the resulting
bridge oscillation.  Simultaneous detection of signals from
multiple planes is possible if the different planes to be measured
are driven at distinct modulation frequencies $\omega_{m}$.

\begin{figure}
\begin{center}
\epsfysize=1.7in \epsfbox{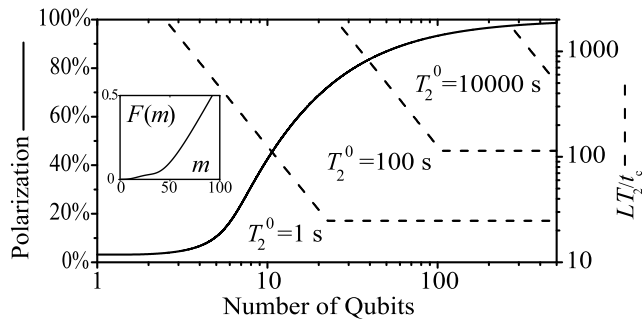}
\end{center}
\caption{A plot showing the scalability of the present scheme. The
solid curve, corresponding to the left axis, shows the
polarization $p$ needed in order for a number of qubits $n$ to be
measurable.  The dashed lines, corresponding to the right axis,
plot the number of logic gates (decoherence time $T_2$ divided by
the pulse sequence cycle time $t_{\text{c}}$) times the length of
a ``block" of the decoupling sequence $L$ against $n$ for several
values of $T_2^0.$ The inset, showing $F(m)$, is explained in the
text. \label{scalingfig}}
\end{figure}

The force resolution for MRFM is limited by thermal fluctuations
of the mechanical oscillator \cite{gabrielson93}. Force
resolutions of $5.6\times10^{-18}$ N/$\sqrt{\text{Hz}}$ have been
reported for single crystal silicon cantilevers at 4 K
\cite{yasamura2000}. The thermal noise threshold of the bridge
structure in Fig. \ref{structure} is estimated to be $\sim
1.2\times 10^{-17} \ \text{N}/\sqrt{\text{Hz}}$ based on a lumped
harmonic oscillator model and assuming a modest quality factor $Q$
of $10^{4}$
\cite{Q_factor}. This model yields an estimated spring constant of
$k \approx 0.0042$~N/m and a resonance frequency
$\omega_{\text{c}}/2\pi \approx 23$~kHz. The detectable signal
will depend upon the initial polarization $p$ after optical
pumping and algorithmic cooling. The force from the subensemble
magnetization corresponding to a pseudo-pure state is estimated
\cite{warren97,ppsnote} as
\begin{equation}
\label{SNR}
    F^z=\frac{\hbar \Delta\omega}{2a} N
    \left[\left(\frac{1+p}{2}\right)^n-\left(\frac{1-p}{2}\right)^n\right].
\end{equation}
The number of qubits available in this scheme may be found by
maximizing $n$ in Eq. (\ref{SNR}) such that the force exceeds the
thermal noise threshold; the results of such maximization are
shown in Fig. \ref{scalingfig}.  At low $p$, exponential
improvements in $p$ are needed to increase the number of
measurable qubits $n$. Once $p$ exceeds about 60\%, however, we
find $n\sim (1+p)/(1-p)$,
escaping the exponential downscaling which plagues solution NMR.

There are several possible sources of decoherence in this
proposal: magnetic fluctuations in the dysprosium, thermal
currents in the dysprosium, fluctuations of paramagnetic
impurities in the silicon, thermal motion of the bridge, and
uncontrolled dipolar couplings between nuclei. We estimate that
the latter two sources are the most important, so we limit our
present discussion to them.

A calculation using the thermal noise statistics of a
high-sensitivity mechanical oscillator \cite{gabrielson93}
estimates the $T_2$ timescale due to the bridge's thermal drift as
$T_2^{\text{c}}=k \omega_{\text{c}}Qa^2/\Delta\omega^2
k_{\text{B}} T\approx 25$~s.
Active feedback stabilization is
expected to increase this timescale by as many as four orders of
magnitude, bringing it close to the lengthy $T_1$ timescale.


There are two sorts of uncontrolled dipolar couplings to consider.
There are the in-plane couplings, in which the participating
nuclei have equal Larmor frequencies and can thus participate in
spin flip-flop processes.  Since the average distance between
chains is $\sim$15~nm, these processes cause decoherence on a time
scale of order $T_2^{\text h}\sim
4\pi(15~\text{nm})^3/\gamma^2\hbar\mu_0 \sim 100$~s.  This already
lengthy timescale may be increased by the addition of dipolar
refocussing sequences such as WAHUHA \cite{haeberlin}.

There will also be spurious couplings between nuclei in one
homogeneous plane and copies of nuclei in the next. Such couplings
occur when two qubits are recoupled for a logic gate, and they
cause a small error in each logic gate. To estimate this error,
suppose we recouple one qubit to another qubit $m$ planes away.
The strongest coupling seen by a nucleus is the in-chain coupling,
and its rate is $\sim \delta\omega/m^3$. The couplings to all of
the neighbors may be treated as a $T_2$ decoherence process, with
\begin{equation}
\left(\frac{1}{T_{2m}^{\text{r}}}\right)^2
=\frac{1}{16}\left(\frac{\delta\omega}{m^3}\right)^2
\sum_i\frac{\left(\lambda_i^2/m^2-2\right)^2}{\left(\lambda_i^2/m^2+1\right)^5}.
\end{equation}
Here, $T_2$ has been estimated as the inverse square root of the
second moment \cite{haeberlin} and $\lambda_i$ is the lateral
distance to the $i$th chain normalized by $a$.  The ratio of the
approximate gate time $m^3/\delta\omega$ to $T_{2m}^{\text{r}}$ is
plotted in the inset of Fig. \ref{scalingfig}. This function,
$F(m)$, represents the approximate error in the gate when
attempting to couple qubits $m$ planes apart.  To keep gate errors
low, the computer should couple only nearby neighbors and handle
more distant couplings by bit swapping.  In this way, gate errors
due to unrefocussed nuclear couplings between chains is limited to
approximately $F(1)\approx 10^{-6}.$

These cross-couplings also influence the clock speed of this
computer, as we now explain.  When qubits are decoupled by the
Hadamard scheme mentioned above, the resulting pulse sequence has
a clock time $t_{\text{c}}=Ln^2/\Delta\omega$, where $n$ is the
number of qubits being decoupled and $L/\Delta\omega$ is the
amount of time devoted to one $\pi$ pulse \cite{leung99}. As
qubits are added, the amount of time needed to recouple them
grows. For very large $n$, however, some qubits in the chain
become so distant that the $r_{ij}^3$ factor in
Eq.~(\ref{ijcouple}) renders their interaction negligibly small.
In this case the Hadamard pulse scheme can be truncated to
decouple qubits only in sets of $l$, and a new decoherence process
is introduced with $T_{2l}^{\text{t}}\delta\omega=l^3
[1+F^2(l)]^{-1/2}.$ If we compare the total decoherence time
$T_2=(1/T_2^0+1/T_{2l}^{\text{t}})^{-1}$, where $T_2^0$ combines
all other decoherence processes, to the clock speed
$t_{\text{c}}$, we find that there is an optimum $l$ at which to
truncate the Hadamard pulse scheme. The effective number of logic
gates $T_2/t_{\text{c}}$, therefore, at first decreases as $n^2$
and then flattens once $n$ reaches this optimum $l$, as shown in
Fig. \ref{scalingfig}.

By consulting Fig. \ref{scalingfig}, we see that this scheme can,
for sufficiently large $T_2^0$ and sufficiently high
polarizations, allow substantially more qubits than solution NMR,
without the need for single-spin measurement or unrealistic
advances in fabrication, measurement, or control technologies.

The work at Stanford was partially supported by NTT Basic Research
Laboratories.  T. D. L. was supported by the Fannie and John Hertz
Foundation.  The work at Keio was partially supported by
Grant-in-Aid for Scientific Research from JSPS and the KAST
Research Grant. We would like to thank fruitful discussions with
A. Verhulst, A. D\^ana, T. Ishikawa, O. D. Dubon, and Y. Saito.


\begin{references}
\bibitem{bmrvv2001note}
    Alternatively, the ``hot" qubits could be recooled if their
    $T_1$ can be made substantially less than the other qubits.
    This allows the restrictive entropy limit to be escaped \cite{bmrvv2001}.
    Such rapid cooling may
    be possible in this proposal if the optical pumping field can
    be limited to a very small ($\sim$100~nm) depth from the
    silicon surface.
\end{references}
\end{document}